\documentclass[letterpaper, 10 pt, conference]{ieeeconf}  

\IEEEoverridecommandlockouts                              
\usepackage{xcolor}
\usepackage{soul}
\usepackage{graphics} 
\usepackage{graphicx}
\usepackage{float}
\usepackage{cleveref}
\usepackage{url}
\usepackage{amsmath} 
\usepackage{amssymb}  
\usepackage{algorithm}
\usepackage[noend]{algpseudocode}

\DeclareMathOperator*{\diag}{diag}

\usepackage{cite}
\usepackage[keeplastbox]{flushend}
\usepackage{booktabs}

\usepackage{pgfplots}
\usepackage{pgfplotstable}
\usepackage{tikz}

\pgfplotsset{compat=newest}
\usepgfplotslibrary{groupplots}
\pgfplotsset{every axis legend/.append style={legend cell align=left}}
\pgfplotsset{every axis/.append style={font=\footnotesize}}

\title{\LARGE \bf
Simultaneous active parameter estimation and control using sampling-based Bayesian reinforcement learning
}

\author{Patrick Slade$^{1}$, Preston Culbertson$^{1}$, Zachary Sunberg$^{2}$, and Mykel Kochenderfer$^{2}$%
\thanks{This work is supported by the National Science Foundation Graduate Research Fellowship Program Grant DGE-1656518 and the Stanford Graduate Fellowship. Toyota Research Institute ("TRI")  provided funds to assist the authors with their research but this article solely reflects the opinions and conclusions of its authors and not TRI or any other Toyota entity.}%
\thanks{$^{1}$P. Slade and P. Culbertson are with the Department of Mechanical Engineering, Stanford University, Stanford, CA 94305 USA (e-mail: \{patslade, pculbertson\}@stanford.edu).}%
\thanks{$^{2}$Z. Sunberg and M. Kochenderfer are with the Department of Aeronautics and Astronautics, Stanford University (e-mail: \{zsunberg, mykel\}@stanford.edu).}%
}
\begin{document}

\maketitle
\thispagestyle{empty}
\pagestyle{empty}

\begin{abstract}

Robots performing manipulation tasks must operate under uncertainty about both their pose and the dynamics of the system. In order to remain robust to modeling error and shifts in payload dynamics, agents must simultaneously perform estimation and control tasks. However, the optimal estimation actions are often not the optimal actions for accomplishing the control tasks, and thus agents trade between exploration and exploitation. This work frames the problem as a Bayes-adaptive Markov decision process and solves it online using Monte Carlo tree search and an extended Kalman filter to handle Gaussian process noise and parameter uncertainty in a continuous space. MCTS selects control actions to reduce model uncertainty and reach the goal state nearly optimally. Certainty equivalent model predictive control is used as a benchmark to compare performance in simulations with varying process noise and parameter uncertainty.

\end{abstract}

\section{INTRODUCTION}
\subsection{Motivation}

Flexibility and robustness in real-world robotic systems is essential to making them human-friendly and safe. Planning for important tasks in robotics, such as localization and manipulation, requires an accurate model of the dynamics \cite{lavalle2001randomized,janson2015fast,thrun2005probabilistic}. However, the dynamics are often only partially known. For example, order-fulfillment robots move containers with varying loads in warehouses  \cite{d2008future}, autonomous vehicles traverse variable terrain and interact with human drivers \cite{beiker2012legal, sadigh2016gathering, sunberg2017internal}, and nursing robotic systems use tools and interact with people \cite{pineau2003towards}. Payload shifts, environment conditions, and human decisions can dramatically alter the system dynamics. Even for motion determined by physical laws, there are often unknown parameters like friction and inertial properties. 

In these cases, the robot must estimate the system dynamics to achieve its goals. Unfortunately, this estimation task often conflicts with the original goal task. The robot must balance \emph{exploration} to gain knowledge of the dynamics and \emph{exploitation} of its current knowledge to obtain rewards.

\subsection{Related Work}

Often this is solved with certainty equivalent control where a robot plans assuming an exact  dynamics model\cite{bertsekas1995dynamic}. This is usually the most likely or mean dynamics model. Typically, this dynamics model will update every time it receives an observation and make a new plan up to a fixed horizon that is optimal for the updated dynamics model. This approach is known as model predictive control (MPC) \cite{bertsekas1995dynamic,garcia1989model}. In some cases, performance can improve if a range of dynamic models are considered. This approach is known as robust MPC \cite{campo1987robust,allwright1992linear,kouvaritakis2015model}. These approaches work well in many cases, however they do not encourage exploration, and are thus suboptimal for some problems \cite{kaelbling1998planning}.

There have been a variety of principled approaches to handle the exploration-exploitation trade-off. One body of research has referred to this challenge as the ``dual control'' problem \cite{feldbaum1960dual}. It was shown that an optimal solution could be found using dynamic programming (DP). However, since the state, action, and belief spaces are continuous, the exact solution is generally intractable. Approximate solutions include a variety of methods such as adaptive control, sliding mode control, and stochastic optimal control \cite{filatov2000survey}. Adaptive controllers typically first perform an estimation task to reveal unknown parameters and then perform the control task. This is suboptimal, the agent could potentially use the estimation actions to also begin performing the control task.

Another popular approach is reinforcement learning (RL) \cite{Wiering2012}. In RL, the underlying planning problem is a discrete or continuous Markov decision process (MDP) with unknown transition probability distributions. RL agents interact with the environment to accrue rewards and may learn the transition probabilities along the way if it helps in this task. If the prior distribution of these transition probabilities is known, the policy that will collect the most reward in expectation, optimally balancing exploration and exploitation, may be calculated by solving a Bayes-adaptive Markov decision process (BAMDP) \cite{kochenderfer2015decision}. Unfortunately BAMDPs are, in general, computationally intractable, so approximate methods are used \cite{guez2013scalable}. Since any BAMDP can be recast as a partially observable Markov decision process (POMDP), similar approximate solution methods are used. Both offline \cite{bai2013planning} and online \cite{guez2013scalable} POMDP methods have been used for RL in discrete state spaces, and others \cite{chen2016pomdplite} are easily adapted.

There has been limited work on continuous spaces. Value iteration has been used for problems with Gaussian uncertainty \cite{webb2014online}. Controls are locally optimized along a pre-computed trajectory as an extended Kalman filter (EKF) updates the belief over the model parameters. Exploration is encouraged by heuristically penalizing model uncertainty in the reward function. The control designer chooses how much to explore rather than the algorithm optimally calculating it.

\subsection{Contribution}

This research applies Monte Carlo tree search (MCTS) to solve a close approximation to the BAMDP for a problem with continuous state, action, and observation spaces, an arbitrary reward function, Gaussian process noise, and Gaussian uncertainty in the model parameters. An EKF updates the belief of the unknown parameters and double progressive widening (DPW) \cite{couetoux2011continuous} guides the expansion of the tree in continuous spaces. \Cref{sec:background} gives an introduction to the problems and methods considered, \Cref{sec:problem,sec:approach} give detailed descriptions of the problem and approach, and \Cref{sec:results} shows simulations comparing MCTS with a certainty-equivalent MPC benchmark for two problems.

\section{Background} \label{sec:background}
This section reviews techniques underlying our work: sequential decision making models, MCTS, and EKFs.
\subsection{MDPs, POMDPs, BAMDPs}
	A Markov decision process is a mathematical framework for a sequential decision process in which an agent will move, typically stochastically, between states over time, accruing various rewards for entering certain states, but may affect their trajectory and rewards by taking actions at each time step. A MDP is defined by the tuple $(\mathcal{S},\mathcal{A},T,R,\gamma)$, where:
\begin{itemize}
\item $\mathcal{S}$ is the set of states the system may reach,
\item $\mathcal{A}$ is the set of actions the agent may take,
\item $T(s'\mid s,a)$ is the probability of transitioning to state $s'$ by taking action $a$ at state $s$,
\item $R(s,a)$ is the reward (or cost) of taking action $a$ at state $s$, and
\item $\gamma \in [0,1]$ is the discount factor for future rewards.
\end{itemize}

Solving an MDP develops a policy, $\pi(s):\mathcal{S} \rightarrow \mathcal{A}$, which maps each state to an optimal action that will accrue the most rewards in expectation over some planning horizon.

For infinite-horizon, discounted MDPs, the optimal policy has been shown to satisfy the Bellman equation \cite{bertsekas1995dynamic}.
For small, discrete state and action spaces, the Bellman equation may be solved explicitly with DP. However, for problems with large or continuous state and action spaces approximate solutions use approximate DP (ADP) \cite{powell2011}. Next, we discuss MCTS, the online ADP algorithm applied in this work.

A partially observable MDP (POMDP) is an extension of an MDP, where the agent cannot directly observe its true state, only receiving observations which are stochastically dependent on this state. A POMDP thus contains a model of the probability of seeing a certain observation at a specific state. The agent forms a belief state, $b$ which encodes the probability of being in each state $s$. The agent updates its belief at each step, depending on its previous action, the reward received, and the observation it took. Since the agent may hold any combination of beliefs about its location in the state space, the belief space $\mathcal{B}$ is continuous, making POMDPs computationally intensive to solve.

A Bayes-adaptive MDP (BAMDP) has transition probabilities that are only partially known. That is, the decision-making agent initially only knows a prior distribution of the transition probabilities. As the agent interacts with the environment, it extracts information of the transition probabilities from the history of states and actions that it has visited. A BAMDP becomes a POMDP by augmenting the state with the unknown parameters defining the transition probabilities.

A POMDP is actually an MDP where the state space is the belief space of the original POMDP \cite{kaelbling1998planning}, sometimes called a belief MDP. The transition dynamics of the belief MDP are defined by a Bayesian update of the belief when an action is taken and an observation received. Not all belief MDPs are POMDPs; in a POMDP, the reward function is specified only in terms of the POMDP state and action and is not a general function of the belief, so penalties cannot be explicitly assigned to uncertainty in the belief. When the belief update is computationally tractable, ADP techniques designed for MDPs may be applied to POMDPs by applying them to the corresponding belief MDP. 

\subsection{Monte Carlo Tree Search}

MCTS is a sampling-based online approach for approximately solving MDPs. MCTS uses a generative model $G$ to get a random state and reward $(s',r) \sim G(s,a)$. It performs a forward search through the state space, using $G$ to draw prospective trajectories and rewards. In MCTS, a tree is created with alternating states and actions, with a roll-out performed to select a policy \cite{browne2012survey}.
In general, MCTS involves stages:
selection, expansion, rollout, and propagation \cite{bertsimas2014comparison}.

\subsection{Extended Kalman Filter}
	In problems with linear Gaussian dynamics and observation functions, it has been shown that the optimal observer is a Kalman filter. A Kalman filter is an iterative algorithm that can exactly update Gaussian beliefs over the state, given the action taken, the observation received, and the transition and observation models \cite{thrun2005probabilistic}.
    
    For systems with nonlinear dynamics, an EKF may be used to approximately update the beliefs with each step \cite{thrun2005probabilistic}. Let $x_k$, $a_k$, and $o_k$ be the state, action, and observation at time $t=k$. For a system with nonlinear-Gaussian dynamics, the transition and observation models can be expressed as
\begin{equation}\label{eq:transition}
x_{k+1} = f(x_{k},a_{k}) + v_k
\end{equation}
\begin{equation}\label{eq:observation}
o_k = h(x_k,a_k) + w_k\text,
\end{equation}
where $f$, $h$ are nonlinear functions and $v$, $w$ are normally distributed. The Gaussian belief has mean $\hat{x}$ and covariance $\Sigma$. The EKF updates the belief at each timestep by predicting a new state and covariance based on the action taken, i.e.
\begin{equation}
\label{eq:pstate}
\hat{x}_{k\mid k-1} = f(\hat{x}_{k-1\mid k-1},a_{k-1})
\end{equation}
\begin{equation}
\label{eq:pvar}
\Sigma_{k\mid k-1} = F_{k-1}\Sigma_{k-1 \mid k-1}F_{k-1}^{T} + Q_{k-1}\text,
\end{equation}
where $F$ is the linearized dynamics about the current belief.
\begin{equation}
F_{k-1} = \left. \frac{\partial f}{\partial x} \right |_{\hat{x}_{k-1 \mid k-1},u_{k-1}}
\end{equation}
A residual between the observation and predicted state is:
\begin{equation} \label{eq:residual}
r_k = o_k - h(\hat{x}_{k \mid k-1},a_{k-1})\text.
\end{equation}
The residual covariance, $S_k$ and Kalman gain, $K_k$ are
\begin{equation}
S_k = H_k \Sigma_{k \mid k-1} H_k^T + R_k
\end{equation}
\begin{equation}
K_k = \Sigma_{k \mid k-1}H_k^T S_k^{-1}
\end{equation}
where $H_k$ is the linearized observation function.
\begin{equation}
H_k = \left. \frac{\partial h}{\partial x} \right |_{\hat{x}_{k \mid k-1},u_{k-1}}
\end{equation}
Finally, the belief update is completed according to
\begin{equation}
\label{eq:bsu}
\hat{x}_{k \mid k} = \hat{x}_{k \mid k-1} + K_k r_k 
\end{equation}
\begin{equation}
\label{eq:bvu}
\Sigma_{k \mid k} = (I-K_k H_k)P_{k \mid k-1}\text.
\end{equation}

Due to the linearization of the dynamics and observation models, EKFs will not always converge and are generally not optimal observers. However, in practice they perform  well for most systems where the true belief is unimodal.
\section{Problem Formulation} \label{sec:problem}
Let us consider a robot trying to control a system with linear-Gaussian dynamics. The system is described by
\begin{equation}
\label{eq:basicModel}
x_{k+1} = f(x_k,u_k;p) + v_k\text,
\end{equation}
where $x_k$, $u_k$ are the state and control action at time $t=k$, $p$ is a set of parameters that the system dynamics depend on, the process noise $v_k \sim \mathcal{N}(0,Q)$, and $f$ is a time-invariant function that is linear with respect to $x_k$ and $u_k$,
\begin{equation}
f(x_k,u_k;p) = A(p)x_k + B(p)u_k\text.
\end{equation}

The observation model is described by the linear equation
\begin{equation}
o_k = h(x_k, u_k;p) + w_k\text,
\end{equation}
with observation, $o_k$ taken at time $t=k$, the measurement noise $w_k \sim \mathcal{N}(0,R)$, and $h$ is a time-invariant linear function
\begin{equation}
h(x_k,u_k;p) = C(p)x_k+D(p)u_k\text.
\end{equation}

While $f$ and $h$ are physical equations known \textit{a priori}, the parameters $p$ are not known beforehand. They can be appended to the state vector to form a state-parameter vector,
\begin{equation}
s_k = \left[ \begin{array}{c} x_k \\ p \end{array} \right]\text.
\end{equation}
Thus, the system dynamics for $s_k$ may be described by
\begin{equation}
s_{k+1} = \left[ \begin{array}{cc} A(p) & 0 \\ 0 & I \end{array} \right] s_k + \left[ \begin{array}{c} B(p) \\ 0 \end{array} \right] u_k + \tilde{v}_k\text,
\end{equation}
where $\tilde{v} \sim \mathcal{N}(0,\diag(Q,P))$ and $P$ is a parameter drift matrix.
The observation model is described by
\begin{equation}
\label{eq:obs}
o_k = \left[ \begin{array}{cc} C(p) & 0 \end{array} \right]s_k + D(p)u_k + w_k\text.
\end{equation}


	If we use an EKF to describe the belief about the current state in the state-parameter space, we form a belief MDP over all possible EKF states. This belief MDP is described by the tuple $(\mathcal{S},\mathcal{A},T,R)$, where:
\begin{itemize}
\item $\mathcal{S}$ is the space of all possible beliefs. Since the belief maintained by the EKF is Gaussian, it can be described by the mean and covariance, $b_k = (\hat{s}_k,\Sigma_k)$.
\item $\mathcal{A}$ is all possible actions that the agent may take.
\item $T(b' \mid b,a)$ is a distribution over possible EKF states after a belief update. This distribution depends on the observation model. It is difficult to represent explicitly, so it is implicitly defined by the generative model, $G$.
\item $R(b,a)$ is a reward function for a given belief and action. It is constructed as desired for a given control task. In our work, we approximate $R(b,a) = R(\hat{s}_k,a)$, a linear reward for the estimated mean state and action.
\end{itemize}

The generative model for the belief MDP is
\begin{equation}
b_{k+1} = G(b_k, a_k)
\end{equation}
with $G$ defined by equations (\ref{eq:pstate}), (\ref{eq:pvar}), (\ref{eq:residual}), (\ref{eq:bsu}), and (\ref{eq:bvu}). The observation, $o_k$, used in (\ref{eq:residual}) is a random variable determined by (\ref{eq:transition}) and (\ref{eq:observation}); it is not the most likely observation. Solving this belief MDP gives a policy that optimally maximizes the sum of expected rewards over some planning horizon.

\section{Approach} \label{sec:approach}

This section discusses the use of MCTS and MPC with an EKF to control a system with unknown parameters.

\subsection{MCTS}

Our approach uses the upper confidence tree (UCT)~\cite{browne2012survey} and DPW~\cite{couetoux2011continuous} extensions of MCTS. The tree is built as UCT expands action nodes maximizing an upper confidence estimate
\begin{equation}
UCB(s, u) = \tilde{Q}(s,u) + c \sqrt{\frac{\log N(s)}{N(s,u)}}\text,
\end{equation}
where $\tilde{Q}(s,u)$ estimates the state-action value function from rollout simulations and tree search, $N(s,u)$ counts the times action $u$ is taken from state $s$, with exploration constant, $c$. This balances exploration-exploitation as the tree expands.

DPW defines tree growth for large or continuous state and action spaces. To avoid a shallow search, the number of children of each state-action node ($s, u$) is limited to
\begin{equation}
k N(s,u)^{\alpha}\text,
\end{equation}
where $k$ and $\alpha$ are parameter constants tuned to control the widening of the tree. With an increase in $N(s, u)$ the number of children also grows, widening the tree. The number of actions explored at each state is limited similarly.

\subsection{Model Predictive Control}

MPC is a technique for online calculation of a policy \cite{garcia1989model}. Its extensive use in control system design is due to its ability to explicitly meet state and control constraints. For our implementation of MPC, the optimization problem is constrained by the dynamics, $f(x_k, u_k, p)$, and the maximum control effort, $u_{max}$. At each step a series of control actions to maximize a reward function are found for a fixed horizon from the current state and the first action is taken \cite{bertsekas1995dynamic}.

\subsection{Basic Approach}

A large number of nonlinear stochastic systems can be near-optimally controlled with an EKF and MCTS or MPC. The purpose of this method is to produce and execute a control policy. Any system described by a model in the form of (\ref{eq:basicModel}) with an approximately Gaussian initial belief state can use MCTS or MPC to find a suitable control action.

Taking this action propagates the true state, $s$, which will be partially observed. This observation and action will update the belief state with the EKF, improving our estimates about the parameters. This process is shown in Algorithm \ref{alg1}.

\begin{algorithm}
\begin{algorithmic}
\caption{Simultaneous estimation and control}
\label{alg1}
\Require $b_0(s), s_0$
\For{$t \in [0, T)$}
\State {$u_t \leftarrow \Call{Policy}{b_t}$}
\State{$o_{t} \leftarrow \Call{ReceiveMeasurements}{s_{t}}$}
\State{$b_{t+1} \leftarrow \Call{EKF}{b_t,u_t,o_{t}}$}
\State{$s_{t+1} \leftarrow \Call{PropogateDynamics}{s_t,u_t}$}
\EndFor
\Return $u_i|_{i=0}^{T}$ 
\end{algorithmic}
\end{algorithm}
\vspace{-5mm}

\subsection{Implementation Details}

Measurement noise was removed from the tests so the effects of process noise and estimated parameter uncertainty were isolated. The magnitude of control inputs was limited to $u_{max}$ (see Table \ref{tab:params}). Filtering the control signal is recommended for experimental validation to not damage actuators.

A linear reward function was implemented with a weighted L1 norm penalty for the position, speed, and control effort
\begin{equation}
R_{L1}(x_k,u_k) = \left[\begin{array}{cc} R_{pos} & 0 \\ 0 & R_{vel} \end{array}\right] x_k + R_u u_k \text.
\end{equation}
The values for $R_{pos}$, $R_{vel}$, and $R_u$ are given in Table \ref{tab:params}. For performance comparison purposes a quadratic reward function is used with a L2 norm penalty for the same terms
\begin{equation}
R_{L2}(x_k,u_k) = x_k^T \left[\begin{array}{cc} R_{pos} & 0 \\ 0 & R_{vel} \end{array}\right] x_k + u_k^T R_u u_k\text.
\end{equation}
Real-world constraints were included as a minimum threshold for physical values such as mass, friction, and inertia so the dynamic models held. The MCTS rollout used a position controller to select a force proportional to the distance from the goal state. The MCTS with DPW implementation is from the POMDPs.jl package \cite{egorov2017pomdps}. The optimization in the MPC controller was solved with the Convex.jl package \cite{udell2014convex}.
    
\section{Simulation and Results} \label{sec:results}

Models of a 1D double integrator and robot performing planar manipulation tested our MCTS approach against MPC, a certainty equivalent optimal control benchmark.

\subsection{1D Double Integrator Model}

We first considered the control problem where an agent applies force to a point mass in one dimension with dynamics
\begin{equation}
	x_{k+1} = \left[ \begin{array}{cc} 1 & 0 \\ \Delta t & 1 \end{array} \right] x_k + \left[ \begin{array}{c} \frac{\Delta t}{m} \\ \left( \frac{\Delta t}{m}\right)^2  \end{array} \right] f_k + v_k
\end{equation}
\begin{equation}
	o_k = \left[ \begin{array}{cc} 1 & 0 \\ 0 & 1 \end{array} \right]x_k + w_k\text.
\end{equation}
Measurement noise is 0 to reduce the number of parameters.

\begin{figure}[t]
\centering
\input{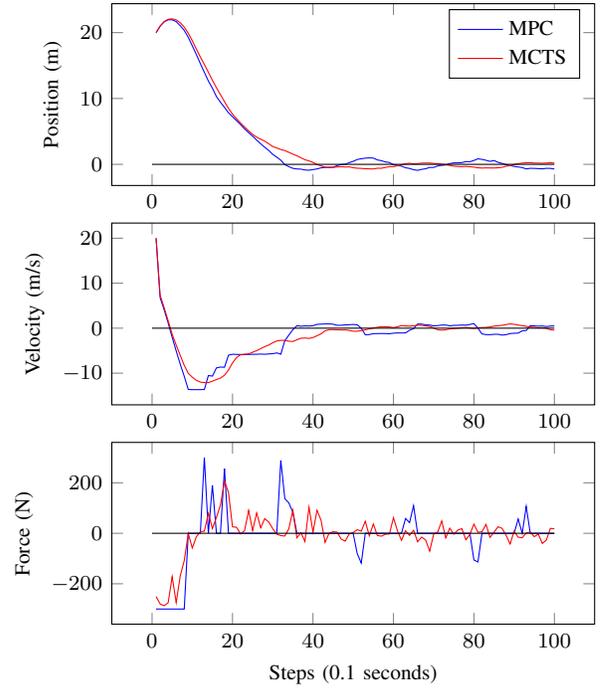}
\vspace{-3mm}
\caption{Position, velocity, and force profiles of the controllers for the 1D double integrator model}
\vspace{-3mm}
\label{fig:profile}
\end{figure}

Fig. \ref{fig:profile} gives physical intuition to the 1D double integrator model. The point mass starts at a given position and velocity with a goal state at the origin. The state profiles for MCTS appear smoother than MPC. This is due to MPC applying large forces for short duration caused by the large penalty an L1 reward function gives small errors.

\subsection{Planar Manipulation (PM) Model}
\begin{figure}[ht]
\centering
\includegraphics[width=0.4\linewidth]{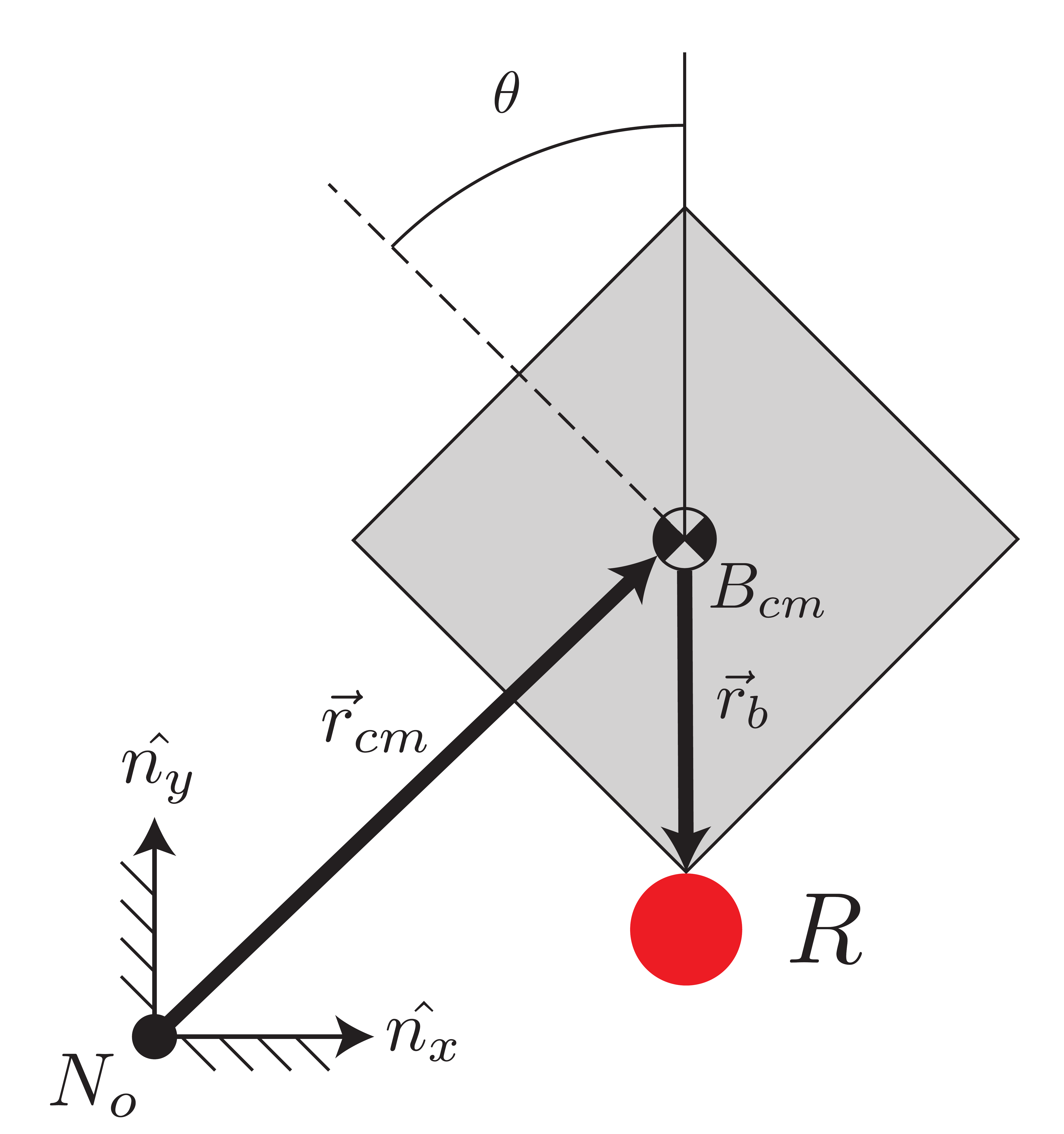}
\vspace{-6mm}
\caption{Schematic of planar  manipulation (PM) task}
\label{fig:2Dmanip}
\end{figure}

We then consider a robot $R$ pushing a box $B$ in the plane, where the agent may apply an arbitrary force in the $x$- and $y$-directions, in addition to a torque. This problem, with its related quantities, is illustrated in Figure \ref{fig:2Dmanip}.

We can describe the dynamics of the system about its center of mass, $B_{cm}$. These are given by
\begin{equation}
\sum \vec{F}_B = \vec{F} - \mu_v \vec{v} =  m \vec{a}
\label{force_eq}
\end{equation}
\begin{equation}
\sum \vec{T}_B= \vec{T} + \vec{r} \times \vec{F} = J \vec{\alpha}\text,
\label{torque_eq}
\end{equation}
where $\mu_v$ is linear friction. The system in discrete-time is
\begin{eqnarray}
	v_{k+1} = a_k  \Delta t+ v_k\\
	p_{k+1} = v_k \Delta t + p_k\text,
\end{eqnarray}
where $\Delta t$ is the discretization step. The system can be put in state-space form with the state vector $x_k = \left[ \begin{array}{cccccc} p_{x,k}, & p_{y,k}, & \theta_k, & v_{x,k}, & v_{y,k}, & \omega_k \end{array} \right]^T$, corresponding to the linear and angular positions and velocities of $B$ in the global frame $N$. Using (\ref{force_eq}) and (\ref{torque_eq}) and an explicit-time integration
\begin{equation}
x_{k+1} = f(x_k,u_k) = A x_k \\ + B(\theta) u_k + v_k\text,
\end{equation}
where
\begin{equation*}
A = \left[ \begin{array}{c|c}
\mathbb{I}^{3x3} & \mathbb{I}^{3x3} \Delta t  \\
\hline \\ [-0.8em]
\huge 0^{3x3}  & \mathbb{I}^{3x3}
\end{array} \right]\text, \end{equation*}
\begin{equation*}
B=  \left[\begin{array}{ccc}
\multicolumn{3}{c}{\Huge 0^{3x3}} \\
\hline \\[-0.8em] \frac{\Delta t}{m} & 0 & 0\\
0 & \frac{\Delta t}{m} & 0\\
B^{3,1} & B^{3,2} & \frac{\Delta t}{J} \end{array}\right]\text,
\end{equation*}
\begin{equation*}
B^{3,1} =  \frac{\Delta t}{J}(\cos(\theta)r_y + \sin(\theta)r_x)\text,
\end{equation*}
\begin{equation*}
B^{3,2} =  \frac{\Delta t}{J}(\cos(\theta)r_x - \sin(\theta)r_y)\text,
\end{equation*}
\begin{equation*}
u_k = \left[ \begin{array}{c}
F_{x,k} \\ F_{y,k} \\ T_k \end{array}\right]\text,
\end{equation*}
where $m$ is the mass of $B$ and $J$ is $B$'s moment of inertia.

For a robot with noisy sensors which measure its position, velocity, and acceleration in $N$, the observation model is
\begin{equation}
	y_k = h(x_k,u_k) + w_k\text,
\end{equation}
where $y_k = \left[ p_{x,k},  p_{y,k},  \theta_k,  v_{x,k},  v_{y,k},  \omega_k,  a_{x,k},  a_{y,k},  \alpha_k \right]^T$.

The measurement functions are given by
\begin{align}
\vec{p} &= \vec{r}_{cm} + \vec{r}_b \\
\vec{v} &= \vec{v}_{cm} + \vec{\omega} \times \vec{r}_b\\
\vec{\alpha} &= \frac{\sum \vec{T}_B}{J} \\
\vec{a} &= \frac{\sum \vec{F_B}}{m}+ \vec{\alpha} \times \vec{r}_b + \vec{\omega}\times(\vec{\omega} \times \vec{r}_b)\text.
\end{align}

This model uses the same stationary goal at the origin with an orientation of 0 degrees. It is given an initial position and orientation with Cartesian and angular velocities.

\subsection{Results}

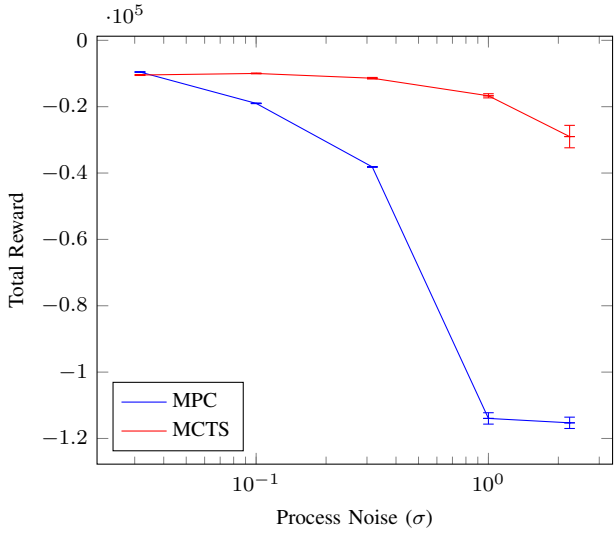
\begin{figure}
\centering
\begin{tikzpicture}[]
\begin{axis}[legend pos = {south west}, xlabel = {Process Noise ($\sigma$)}, xmode = {log}, ylabel = {Total Reward}]\addplot+ [
mark = {none}, blue, error bars/.cd, 
x dir=both, x explicit, y dir=both, y explicit]
table [
x error plus=ex+, x error minus=ex-, y error plus=ey+, y error minus=ey-
] {
x y ex+ ex- ey+ ey-
0.03162277660168379 -9551.87472783024 0.0 0.0 42.88768478123474 42.88768478123474
0.1 -19007.11246279398 0.0 0.0 74.280460532881 74.280460532881
0.31622776601683794 -38151.7739938339 0.0 0.0 93.78848226172849 93.78848226172849
1.0 -113957.57675972211 0.0 0.0 1716.6503868257903 1716.6503868257903
2.23606797749979 -115284.70139880208 0.0 0.0 1708.3164531045459 1708.3164531045459
};
\addlegendentry{MPC}
\addplot+ [
mark = {none}, red, error bars/.cd, 
x dir=both, x explicit, y dir=both, y explicit]
table [
x error plus=ex+, x error minus=ex-, y error plus=ey+, y error minus=ey-
] {
x y ex+ ex- ey+ ey-
0.03162277660168379 -10453.080638804957 0.0 0.0 145.81876969463084 145.81876969463084
0.1 -9985.515249151953 0.0 0.0 105.98397205809113 105.98397205809113
0.31622776601683794 -11419.56974088077 0.0 0.0 236.28774613801812 236.28774613801812
1.0 -16717.60842262585 0.0 0.0 628.0708299908868 628.0708299908868
2.23606797749979 -29008.851639017015 0.0 0.0 3381.869392185954 3381.869392185954
};
\addlegendentry{MCTS}
\end{axis}

\end{tikzpicture}
\caption{Effects of different levels of process noise on reward for the 1D double integrator model}
\vspace{-2mm}
\label{fig:process1D}
\end{figure}

The simulations performed on both models compare the rewards for MCTS and MPC for two varied parameters: process noise and initial parameter uncertainty. Fig. \ref{fig:process1D} compares rewards of MCTS and MPC for the 1D double integrator model while varying process noise. The simulation parameters are detailed in Table \ref{tab:params}. While MPC and MCTS perform similarly for small process noise, the reward accrued by MPC decreases significantly faster as process noise increases. Process noise greater than 1.0 has little impact on the reward, indicating saturation when the process noise is so large the policy cannot reach the goal region. 30 trials were run for each case with standard error of the mean indicated by the error bars. 

The standard deviation of the initial parameter estimate was then varied for a constant process noise with variance 3.0. This process noise level was chosen because it highlights the rapid change in reward caused by a poorer initial mass estimate. Fig. \ref{fig:mass1D} shows the MCTS reward mostly unaffected by this uncertainty, while the MPC reward decreases rapidly. This highlights the advantage of exploration-exploitation from MCTS. Searching for actions using multiple estimates of the mass and moving towards the goal allows it to get better knowledge of the system and increase reward. MPC takes the first action of an open-loop certainty-equivalent plan recalculated every step.

\begin{figure}
\centering
\begin{tikzpicture}[]
\begin{groupplot}[group style={horizontal sep = 1.0cm, vertical sep = 0.0cm, group size=3 by 1}]
\nextgroupplot [width = {4cm}, ymax = {0.0}, height = {8cm}, xlabel = {L1}, ylabel = {Total Reward}, xmax = {2.5}, xmin = {0.5}]\addplot+ [
mark = {none}, ybar,fill=blue, error bars/.cd, 
error bar style = {black, very thick}, x dir=both, x explicit, y dir=both, y explicit]
table [
x error plus=ex+, x error minus=ex-, y error plus=ey+, y error minus=ey-
] {
x y ex+ ex- ey+ ey-
1.0 -113957.57675972211 0.0 0.0 1716.6503868257903 1716.6503868257903
};
\addplot+ [
mark = {none}, ybar,fill=red, error bars/.cd, 
error bar style = {black, very thick}, x dir=both, x explicit, y dir=both, y explicit]
table [
x error plus=ex+, x error minus=ex-, y error plus=ey+, y error minus=ey-
] {
x y ex+ ex- ey+ ey-
2.0 -16717.60842262585 0.0 0.0 628.0708299908868 628.0708299908868
};
\nextgroupplot [width = {4cm}, legend pos = {south east}, ymax = {0.0}, height = {8cm}, xlabel = {L2}, ymin = {-130000.0}, xmax = {2.5}, xmin = {0.5}, legend style = {thick}]\addplot+ [
mark = {none}, ybar,fill=blue, error bars/.cd, 
error bar style = {black, very thick}, x dir=both, x explicit, y dir=both, y explicit]
table [
x error plus=ex+, x error minus=ex-, y error plus=ey+, y error minus=ey-
] {
x y ex+ ex- ey+ ey-
1.0 -76089.4162521158 0.0 0.0 3309.1970111099904 3309.1970111099904
};
\addlegendentry{MPC}
\addplot+ [
mark = {none}, ybar,fill=red, error bars/.cd, 
error bar style = {black, very thick}, x dir=both, x explicit, y dir=both, y explicit]
table [
x error plus=ex+, x error minus=ex-, y error plus=ey+, y error minus=ey-
] {
x y ex+ ex- ey+ ey-
2.0 -92102.18065129299 0.0 0.0 4802.557174234231 4802.557174234231
};
\addlegendentry{MCTS}
\end{groupplot}

\end{tikzpicture}
\caption{Difference in performance of MCTS and MPC with L1 and L2 reward functions}
\vspace{-4mm}
\label{fig:rewardcomp}
\end{figure}
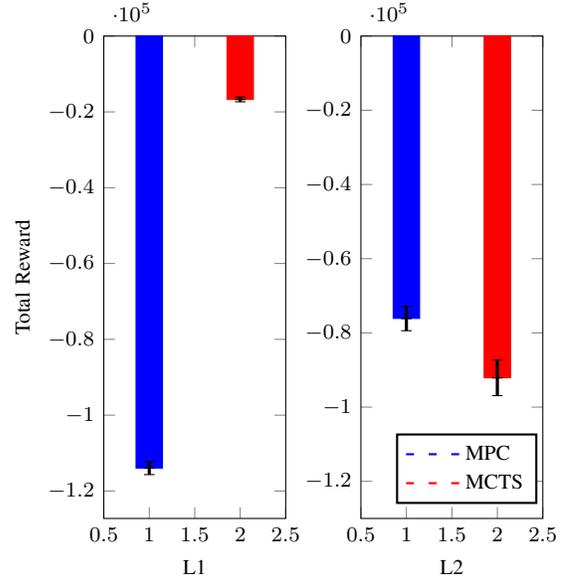

Another interesting topic to consider is the performance of MCTS and MPC considering parameter uncertainty with different reward functions. 
The relative rewards of MCTS and MPC with linear (L1) and quadratic (L2) reward functions are shown in Fig. \ref{fig:rewardcomp}. 
In both cases the process noise had variance 1.0 and the initial mass estimate had variance 10.0.
While MCTS performs significantly better with an L1 reward, it performs slightly worse with an L2 reward.
These results indicate that, with an L2 reward in these domains, taking uncertainty into account with MCTS is not beneficial.
This is due to the L2 reward function placing a small penalty on small errors, allowing MPC to perform well without considering parameter uncertainty.
In another study with a quadratic state reward \cite{webb2014online}, the reward function is augmented with a term penalizing parameter uncertainty.
We suspect that, without this explicit stimulus to explore rather than simply exploit, planning algorithms would not need to actively gather information about the parameters to perform well in domains with quadratic reward.
A chief advantage of the POMDP approach is that the solver automatically gathers information necessary to maximize the reward. Explicit penalization of uncertainty should not be necessary if the state reward function is chosen to accurately reflect the  performance requirements.
It should be noted that convex reward functions were used to allow easy comparison with MPC, but there is no need for the reward function used by MCTS to be convex. Designers can use reward functions that precisely prescribe the desired behavior.

\begin{table}
\caption{Simulation parameters\label{tab:params}}
\begin{center}
\begin{tabular}{ @{}lccc@{} } 
 \toprule
 Parameter & Symbol & Value 1D & Value PM \\ 
 \midrule
 
 Time step & $\Delta t$ & 0.1 s & 0.1 s\\ 
 UCT exploration parameter& $c$ & 300 & 100\\  
 DPW linear parameter & $k$ & 8.0 & 8.0\\ 
 DPW exponent parameter & $\alpha$ & 0.2 & 0.2\\ 
 Measurement noise & $w_k$ & 0.0 & 0.0\\
 Reward scaler for force & $R_u$ & $-1.0$ & $-0.1$\\
 Reward scaler for position & $R_{pos}$ & $-10.0$ & $-1.0$\\
 Reward scaler for velocity & $R_{vel}$ & $-3.0$ & $-1.0$\\
 MCTS search depth & & 20 & 20\\ 
 Position roll-out gain & & 4.0 & 8.0\\
 MCTS-DPW iterations per step & $n$ & 2000 & 10, 100\\
 MPC horizon & & 20 & 20\\ 
 Magnitude of force range & $u_{max}$ & 300.0 & 100.0 \\ 
 Steps per trial & & 100 & 100\\
 Trials per test case & & 30 & 30 \\
 Initial parameter estimate variance & & 10.0 & 10.0\\
 Parameter and estimate lower bound & & 1.0 & 1.0\\
 \bottomrule
\end{tabular}
\end{center}
\vspace{-4mm}
\end{table}

\begin{figure}
\centering
\begin{tikzpicture}[]
\begin{axis}[legend pos = {south west}, xlabel = {Initial Mass Estimate ($\sigma$)}, xmode = {log}, ylabel = {Total Reward}]\addplot+ [
mark = {none}, blue, error bars/.cd, 
x dir=both, x explicit, y dir=both, y explicit]
table [
x error plus=ex+, x error minus=ex-, y error plus=ey+, y error minus=ey-
] {
x y ex+ ex- ey+ ey-
0.31622776601683794 -57264.89740708736 0.0 0.0 1666.7530310064585 1666.7530310064585
1.0 -55926.722479924516 0.0 0.0 1256.0796324241226 1256.0796324241226
3.1622776601683795 -115284.70139880208 0.0 0.0 1708.3164531045459 1708.3164531045459
10.0 -173410.79723537713 0.0 0.0 2032.6969620524378 2032.6969620524378
31.622776601683793 -228509.6793783294 0.0 0.0 2262.6582708909305 2262.6582708909305
100.0 -284868.0754303426 0.0 0.0 3011.8802471153294 3011.8802471153294
};
\addlegendentry{MPC}
\addplot+ [
mark = {none}, red, error bars/.cd, 
x dir=both, x explicit, y dir=both, y explicit]
table [
x error plus=ex+, x error minus=ex-, y error plus=ey+, y error minus=ey-
] {
x y ex+ ex- ey+ ey-
0.31622776601683794 -27411.198898144306 0.0 0.0 2264.111312594718 2264.111312594718
1.0 -25069.88881847581 0.0 0.0 1187.910003558361 1187.910003558361
3.1622776601683795 -29008.851639017015 0.0 0.0 3381.869392185954 3381.869392185954
10.0 -27946.61066190684 0.0 0.0 1590.236918038003 1590.236918038003
31.622776601683793 -27056.791338430874 0.0 0.0 2603.0340326586334 2603.0340326586334
100.0 -35173.02153006554 0.0 0.0 7104.86926660782 7104.86926660782
};
\addlegendentry{MCTS}
\end{axis}

\end{tikzpicture}
\caption{Effects of different initial mass estimate standard deviations on reward for the 1D double integrator model}
\vspace{-2mm}
\label{fig:mass1D}
\end{figure}
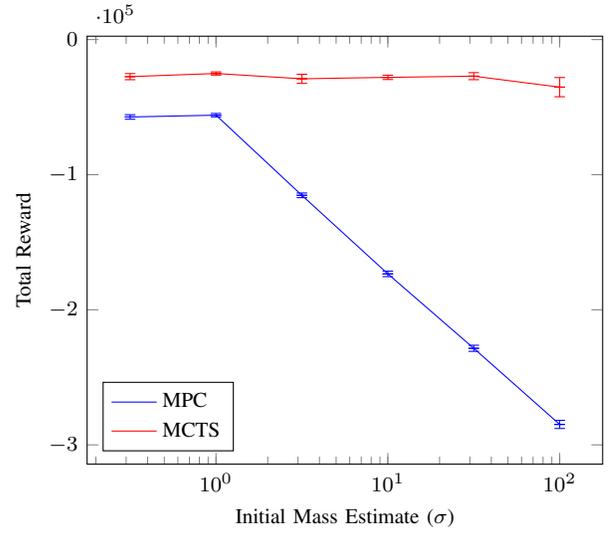

\begin{figure}[t]
\centering
\begin{tikzpicture}[]
\begin{axis}[legend pos = {south west}, xlabel = {Process Noise ($\sigma$)}, xmode = {log}, ylabel = {Total Reward}]\addplot+ [
mark = {none}, blue, error bars/.cd, 
x dir=both, x explicit, y dir=both, y explicit]
table [
x error plus=ex+, x error minus=ex-, y error plus=ey+, y error minus=ey-
] {
x y ex+ ex- ey+ ey-
0.03162277660168379 -7955.161887125055 0.0 0.0 142.83856197540374 142.83856197540374
0.1 -8356.371946255309 0.0 0.0 243.1686309630439 243.1686309630439
0.31622776601683794 -8541.463073885288 0.0 0.0 244.78548981745342 244.78548981745342
1.0 -8941.463073885288 0.0 0.0 244.78548981745342 244.78548981745342
3.1622776601683795 -12116.34873515924 0.0 0.0 311.33304032233343 311.33304032233343
10.0 -24160.455602676157 0.0 0.0 826.7941029540538 826.7941029540538
};
\addlegendentry{MPC}
\addplot+ [
mark = {none}, red, error bars/.cd, 
x dir=both, x explicit, y dir=both, y explicit]
table [
x error plus=ex+, x error minus=ex-, y error plus=ey+, y error minus=ey-
] {
x y ex+ ex- ey+ ey-
0.03162277660168379 -6150.79848372719 0.0 0.0 75.1745592330361 75.1745592330361
0.1 -6250.79848372719 0.0 0.0 75.1745592330361 75.1745592330361
0.31622776601683794 -6250.79848372719 0.0 0.0 95.61288345752789 95.61288345752789
1.0 -6725.901897000962 0.0 0.0 164.63376223164857 164.63376223164857
3.1622776601683795 -9744.706106074487 0.0 0.0 340.8071862050424 340.8071862050424
10.0 -21511.481246242547 0.0 0.0 866.3847836784204 866.3847836784204
};
\addlegendentry{MCTS ($n = 100$)}
\addplot+ [
mark = {none}, green, error bars/.cd, 
x dir=both, x explicit, y dir=both, y explicit]
table [
x error plus=ex+, x error minus=ex-, y error plus=ey+, y error minus=ey-
] {
x y ex+ ex- ey+ ey-
0.03162277660168379 -6096.021619383888 0.0 0.0 94.497321564861 94.497321564861
0.1 -5933.528817137275 0.0 0.0 52.00133381532257 52.00133381532257
0.31622776601683794 -6090.019300199971 0.0 0.0 84.92973893527001 84.92973893527001
1.0 -6627.646860288546 0.0 0.0 161.6738213480106 161.6738213480106
3.1622776601683795 -10014.30812352794 0.0 0.0 414.6793119803321 414.6793119803321
10.0 -23546.97538145727 0.0 0.0 958.2851168286563 958.2851168286563
};
\addlegendentry{MCTS ($n = 10$)}
\end{axis}

\end{tikzpicture}
\caption{Effects of different levels of process noise on reward for the PM model with $n$ representing the number of iterations per step}
\vspace{-2mm}
\label{fig:process2D}
\end{figure}
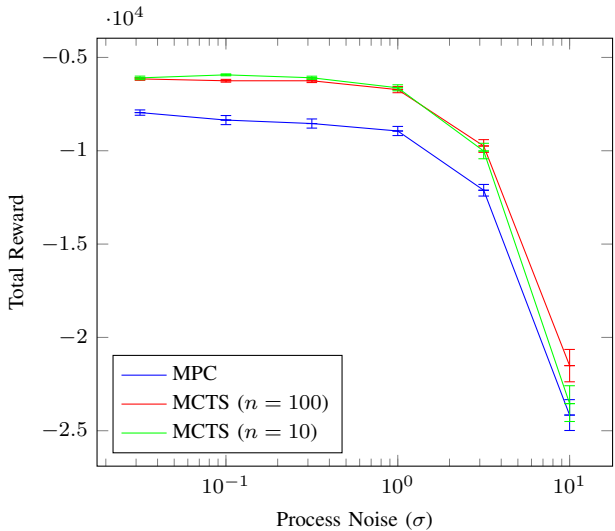

The simulation for varying process noise in the PM model is displayed in Fig. \ref{fig:process2D}. It behaves like the 1D model; MCTS has higher reward for all noise levels. As iterations per step decreases, MCTS tracks closer to MPC. This smaller reward difference than in the 1D case indicates the EKF has trouble reducing uncertainty for five estimated parameters in a non-linear and time-varying system. Thus, rewards decrease with increasing process noise for both controllers.

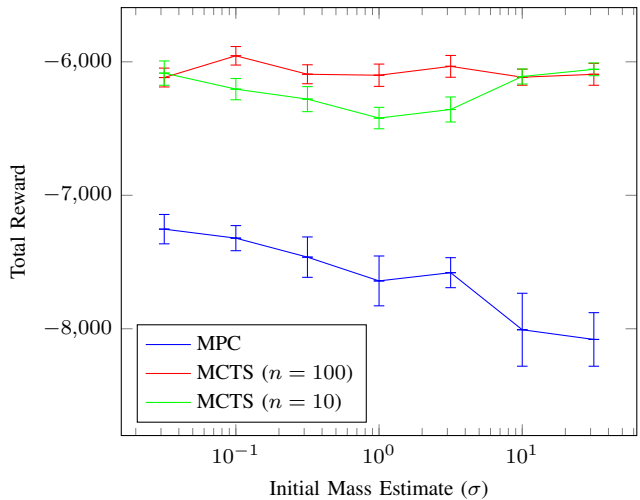
\begin{figure}[t]
\centering
\begin{tikzpicture}[]
\begin{axis}[legend pos = {south west}, xlabel = {Initial Mass Estimate ($\sigma$)}, xmode = {log}, ylabel = {Total Reward}, ymin = {-8800}]\addplot+ [
mark = {none}, blue, error bars/.cd, 
x dir=both, x explicit, y dir=both, y explicit]
table [
x error plus=ex+, x error minus=ex-, y error plus=ey+, y error minus=ey-
] {
x y ex+ ex- ey+ ey-
0.03162277660168379 -7253.662305158 0.0 0.0 109.98206050624525 109.98206050624525
0.1 -7321.074476150321 0.0 0.0 94.11137739838868 94.11137739838868
0.31622776601683794 -7463.314827583352 0.0 0.0 151.2805427918807 151.2805427918807
1.0 -7641.3380297022695 0.0 0.0 186.64680484418136 186.64680484418136
3.1622776601683795 -7579.747758891829 0.0 0.0 112.39570263642686 112.39570263642686
10.0 -8007.2304837441425 0.0 0.0 273.2270092816034 273.2270092816034
31.622776601683793 -8080.029021253357 0.0 0.0 200.7419471090377 200.7419471090377
};
\addlegendentry{MPC}
\addplot+ [
mark = {none}, red, error bars/.cd, 
x dir=both, x explicit, y dir=both, y explicit]
table [
x error plus=ex+, x error minus=ex-, y error plus=ey+, y error minus=ey-
] {
x y ex+ ex- ey+ ey-
0.03162277660168379 -6117.755511299443 0.0 0.0 70.58037624108152 70.58037624108152
0.1 -5954.829430017617 0.0 0.0 69.46133419084848 69.46133419084848
0.31622776601683794 -6093.178178151223 0.0 0.0 70.98927598910393 70.98927598910393
1.0 -6100.511294656724 0.0 0.0 83.56815696715111 83.56815696715111
3.1622776601683795 -6033.943032743673 0.0 0.0 81.61151880241013 81.61151880241013
10.0 -6115.31096960317 0.0 0.0 60.0634568916415 60.0634568916415
31.622776601683793 -6093.642438689464 0.0 0.0 81.9697843817768 81.9697843817768
};
\addlegendentry{MCTS ($n = 100$)}
\addplot+ [
mark = {none}, green, error bars/.cd, 
x dir=both, x explicit, y dir=both, y explicit]
table [
x error plus=ex+, x error minus=ex-, y error plus=ey+, y error minus=ey-
] {
x y ex+ ex- ey+ ey-
0.03162277660168379 -6084.520307960588 0.0 0.0 91.03779325938225 91.03779325938225
0.1 -6204.611476211651 0.0 0.0 79.62670930775698 79.62670930775698
0.31622776601683794 -6279.560380464385 0.0 0.0 93.70241773922798 93.70241773922798
1.0 -6421.411717371161 0.0 0.0 80.26899335628733 80.26899335628733
3.1622776601683795 -6357.105279486244 0.0 0.0 93.84156425790368 93.84156425790368
10.0 -6109.91174772731 0.0 0.0 56.35353922302459 56.35353922302459
31.622776601683793 -6056.17883160397 0.0 0.0 47.957772808318 47.957772808318
};
\addlegendentry{MCTS ($n = 10$)}
\end{axis}

\end{tikzpicture}
\caption{Effects of different initial mass estimate standard deviations on reward for the PM model}
\vspace{-6mm}
\label{fig:mass2D}
\end{figure}

For a constant process noise with standard deviation 0.1, again where MCTS and MPC have a large difference in reward, the standard deviation of the initial parameter estimate is varied for the PM model in Fig. \ref{fig:mass2D}. As the uncertainty of the initial parameters increases, the reward of MPC decreases at a slower rate than seen in the 1D double integrator model. This indicates difficulty stabilizing and reaching the goal for any process noise. The large standard deviations of the initial parameter estimations negatively impact MPC reward while MCTS is unaffected. For real-world tasks like warehouse robots manipulating varying loads, MCTS performs better.

These results indicate MPC's open-loop planning executed in a closed-loop fashion cannot estimate parameters or system propagation as well as MCTS for even simple models with substantial noise. In terms of real-time performance, MCTS using 10 iterations per step ran with an average of 0.156 seconds per step. All simulations were run on an i7 processor. This runtime depends on implementation and hardware, but the same order of magnitude run-time indicates MCTS could be used on systems in real-time. Increasing the number of iterations per step to improve performance when computation time is available provides flexibility onboard.

\section{Conclusions}
This paper posed the control of a robot performing a manipulation task in 1D and 2D for a payload with unknown parameters as a BAMDP. An online, sampling-based approach, MCTS, was used to approximately solve these continuous control problems with Gaussian uncertainty. Empirically, it was shown that this approach effectively balances exploration and exploitation. Even with few samples, MCTS improved parameter estimation and exploitation of the system dynamics to reach a goal state. In simulations with large process and parameter uncertainty, this approach provides policies with significantly higher reward than the commonly used MPC. MCTS algorithms may be a promising way to address estimation and control for real-world autonomous systems.

\section{Acknowledgements}
	
    The authors thank Maxime Bouton for help with the EKF. 
        
\bibliographystyle{ieeetr}
\bibliography{citations}

\end{document}